# Interface Induced Anisotropy and nematic glass/gel state in Jammed Aqueous Laponite Suspensions

A. Shahin and Yogesh M Joshi[*]

Department of Chemical Engineering, Indian Institute of Technology Kanpur 208016, India

S. Anantha Ramakrishna

Department of Physics, Indian Institute of Technology Kanpur 208016, India.

*Corresponding author, E-Mail: joshi@iitk.ac.in

**Abstract**

Aqueous suspensions of Laponite, a system composed of disk-like nanoparticles, are found to develop optical birefringence over several days, well after the suspensions solidified due to jamming. The optical anisotropy is particularly enhanced near the air – Laponite suspension interface over length-scales of several millimetres, which is beyond five orders of magnitude larger than the particle length scale, suggestive of large scale ordering influenced by the interface. The orientational order increases with time and is always greater for higher concentration of salt, higher concentration of Laponite and higher temperature of the suspension. While weakly birefringent, Laponite suspensions covered by paraffin oil do not show any enhancement in the optical anisotropy near the interface compared to that in the bulk. We suggest that the expedited structure formation near the air interface propagating progressively inside the sample is responsible for the observed behaviour. We discuss the observed nematic ordering in the context of glass-like and gel-like microstructure associated with aqueous Laponite suspensions.

## I. Introduction:

Clays are ubiquitous in nature. Their nanoscopic size, anisotropic shape, apparently non-toxic nature and low cost renders them applicable in paper, polymer, petroleum, paint, cosmetic, pharmaceutical, food, etc. industries.[1] Clays have always attracted attention of colloidal community from an academic point of view as well.[2] Their layered nature, charge distribution and richness of microstructure have been a subject of investigation over the last several decades.[3-5] Lately there has been a renewed interest in clays and much of work has been carried out to understand the microstructures and phase behaviours of aqueous clay suspensions.[6-8] Laponite®, a synthetic hectorite clay and its aqueous suspensions have been a subject of intense investigation due to its perplexing phase behaviour and industrial applications.[9] Aqueous suspensions of Laponite are known to show a spectacular increase in viscosity to form a soft solid-like transparent gel/paste that is thermodynamically out of equilibrium. The microstructure that causes very large increase in viscosity, however, is not clearly understood and has been a subject of debate over the last two decades. [10-19]

Laponite, Hydrous sodium lithium magnesium silicate $(Na_{+0.7}[(Si_8Mg_{5.5}Li_{0.3})O_{20}(OH)_4]_{-0.7})$, is a disc shaped nanoparticles with diameter 25 to 30 nm and thickness 1 nm.[9, 20] Particle size distribution of Laponite is fairly monodispersed. Laponite is a smectite (means layer-like) clay and belongs to a family of 2:1 phyllosilicates.[2] A Single layer of Laponite consists of an octahedral magnesia layer sandwiched between two tetrahedral silica layers. Isomorphic substitution of magnesium by lithium within the octahedral layer renders a permanent negative charge to the faces of a particle. The edge of a particle is composed of oxide of silicon, oxides and hydroxides of magnesium (and lithium in very small amount),[10] as suggested by the idealized crystal structure of Laponite.[9] Point of zero charge for MgOH corresponds to pH of 12.5,[21] while that of magnesium oxide is between pH of 10 to 13.[22] Oxide of silicon, on the other hand, is negatively charged beyond pH of 3.5.[22,23] Nonetheless, the crystal structure of Laponite clearly indicates that edge of Laponite must have a negative charge beyond pH of 13. However, it is difficult to establish the point of zero charge associated with the edge precisely as the extremely small size of the particle hinders estimation by a direct measurement. The conductivity measurements of Tawari and coworkers reported positive edge charge for Laponite having pH close to 10.[24] Laponite is hydrophilic and its suspension in water above 1 volume % concentration has a soft solid-like consistency that supports its own weight.[25] Phase behaviour of aqueous suspension of Laponite has been studied and discussed in the literature over the past decade.[10-19] For a Laponite suspension having 1 volume % concentration, mainly two kinds of microstructures have been proposed. The first one is an attractive gel structure formed by a network of particles connected via positive edge and negative face. The second proposed structure is that of a repulsive glass wherein particles are self suspended due to repulsive environment without touching each other. Many groups have observed orientational order in aqueous suspension of Laponite and other clay materials.[17, 26-40] A small angle X-ray scattering study on aqueous Laponite

B suspension for a concentration little above 1 volume % demonstrated existence of nematic like orientational correlations with an order parameter of 0.55 comparable to liquid crystals,[32] suggesting coexistence of weak order along with dynamical defects. Recently a small angle neutron scattering study on Laponite RD suspensions suggested two possibilities: "a long-range orientational order but short-range positional order of Laponite particles" or "short-range order for the repetition of the layers of Laponite discs without any order within each layer".[26] It is, however, important to note that these studies investigate essentially the bulk behaviour.

In this work, we report on Laponite suspensions that develop optical birefringence over many days after going into a completely jammed state. We observe that aqueous suspension of Laponite shows strong enhancement in orientational order in the vicinity of the air/nitrogen interface that extends beyond spatial regions over orders of magnitude greater length-scales than the particle length scale. Remarkably an extent of order near the air interface goes on increasing even though the material is in a physically jammed state.

**II. Material and Methods**

An aqueous dispersion of Laponite RD® (Southern clay products) was prepared by dispersing it in deionised water of pH 10, unless otherwise mentioned, using a turrex drive. The sample was stirred vigorously for 45 minutes to ensure complete dissolution of the clay platelets. The detailed preparation procedure is discussed elsewhere.[10] In this work, we have primarily used 2.8 weight % Laponite suspension with and without salt (NaCl). We have also carried out some experiments in the concentration range of 2 to 3.5 weight % as well as in the pH range of 9 to 13. Concentration of $Na^+$ ions due to externally added NaCl or NaOH is mentioned in mM. Since pH of water is adjusted using NaOH, system with pH 10 without any externally added salt has concentration of $Na^+$ ions around 0.1 mM. This mentioned concentration of $Na^+$ ions, however, does not include the concentration of $Na^+$ ions due to dissociation of counterions associated with Laponite. The freshly prepared samples were filled in the glass cuvettes with square cross section (35 mm × 10 mm × 10 mm). The cuvettes were filled with the sample up to 30 mm of length. In this study we have used two systems, one with Laponite suspension - paraffin oil interface and other with Laponite suspension - air interface. In order to ensure absence of contamination of $CO_2$, we repeated some experiments with $N_2$ atmosphere instead of air. We did not observe any noticeable difference when air was replaced by $N_2$. In this paper, we will therefore use the term air interface to represent both air as well as $N_2$ interface. In oil interface case, the remaining space in cuvette was filled with paraffin oil. In both the cases, the Teflon lid of the cuvette was tightly sealed with an epoxy adhesive in order to prevent any mass transport (paraffin oil, air or water vapour) across the cuvette. This rules out evaporation of water in cuvettes having air interface (it should be noted that the volume of air was less than 15 % of the volume of Laponite suspension). The samples eventually solidify so as to sustain their own weight when inverted. Precaution was taken to prevent any movement of the air or paraffin oil bubble, since the deformation induced by the same may influence the orientation of Laponite particle in the suspension. Some of the Laponite suspension samples were filtered using Millipore Millex-HV 0.45 µm filter before filling the cuvettes.

In a uniaxially anisotropic homogeneous medium, the refractive indices for polarized light with the electric field parallel and perpendicular to the anisotropy axis become different. This results in optical birefringence. Linearly polarized light upon traversing the sample becomes elliptically polarized in general. It can be shown that the intensity of light transmitted across a uniaxial sample placed between two orthogonal polarisers is given by, $I_T = I_0 \sin^2(2\alpha)\sin^2(\Delta n k_0 \delta/2)$,[41] where $I_0$ is the intensity of incident light, $\Delta n$ is the difference of refractive index along the axis and perpendicular to the axis, $k_0$ is a wave vector in the free space $(2\pi/\lambda)$, $\delta$ is the thickness of the sample and $\alpha$ is an angle between the ordinary axis of the anisotropic medium and the polariser axis. Therefore the transmitted intensity is a measure of the birefringence ($\Delta n$).

We conducted the birefringence experiments on aqueous Laponite suspensions wherein the transmittance of light was measured through the sample kept in between crossed polarizers. In these experiments due to large (macroscopic) beam size and the sample thickness, the optical birefringence measured is an average over variously oriented domains. The microscope image as well as the minimal scattering from the sample indicates small domain structures over length-scales of few microns.

A polarizing microscope (Olympus BX -51) was used to investigate the optical transmission properties of the samples at 5X magnification. The axis of the polarizer and analyzer were parallel and perpendicular respectively, to the axis of the cuvette. The spectrum of the transmitted light was measured by an Ocean Optics USB 400 spectrometer connected to the trinocular port of the microscope via an optical fiber. The intensity of the transmitted light was measured as a function of time elapsed since preparation of the sample along the length of a cuvette. Prior to each spectral measurement on the Laponite suspension samples, the intensity of the light through crossed polarizers was measured to account for the fluctuations in the intensity. No resonant spectral features were noticed in the transmittance of light in the wavelength range of 450 nm to 700 nm. We also used a He-Ne laser (632 nm & 1mW) and sheet polarizers to carry out birefringence experiments with a smaller beam size of about 0.6 mm to validate our optical microscope measurements. The intensity beyond the analyzer was measured using a photodiode coupled to an amplifier. A minimal amount of scattering was observed from a path of the laser beam within the sample. The polariser and analyser were simultaneously rotated and the intensity of the transmitted light was also measured as a function of rotation angle.

## III. Results

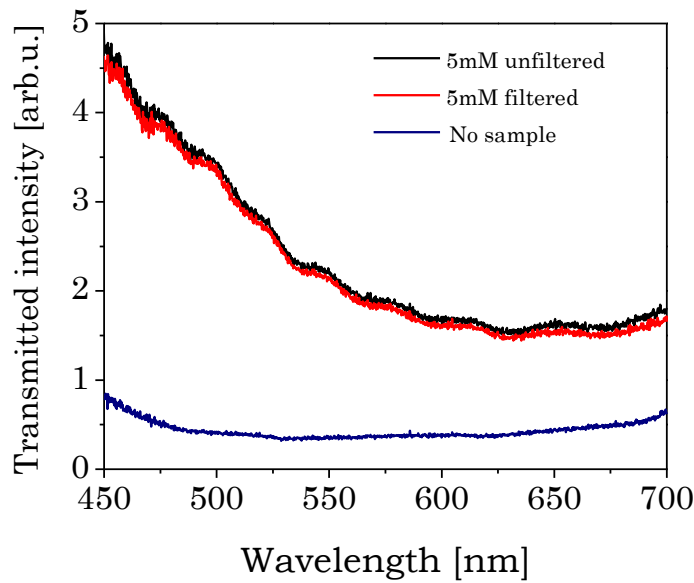

**Figure 1.** Spectrum of the transmitted light through the crossed polarisers with an unfiltered sample (black line) and a filtered sample (red line) of 2.8 weight % Laponite with 5 mM NaCl. Intensity through the crossed polarisers with no sample is also shown in the plot (blue line, bottom curve). The experiments were carried out at a location about 25 mm away from the air interface.

Figure 1 shows a typical spectrum of the transmitted light through crossed polarizers with and without sample (Laponite suspension). It can be seen that intensity of transmitted light gets significantly enhanced due to presence of the sample. We have used two types of samples: suspensions filtered with Millipore Millex-HV 0.45 µm filter and suspensions without any filtration. As shown in figure 1, filtration does not have any noticeable influence on the intensity of the transmitted light. In order to compare the intensity of transmitted light relative to that of crossed polarizer without any sample, we define normalized transmittance as $T_N = \left[\int (I_S/I_{NS}) d\lambda\right] / \left[\int d\lambda\right]$. Here $I_S$ and $I_{NS}$ represent the intensity of transmitted light through crossed polarizer with and without sample respectively. In our case, normalized transmittance $T_N = 1$, therefore represents no birefringence ($I_S = I_{NS}$) suggesting isotropic nature of the sample. On the other hand, $T_N > 1$ represents birefringence ($I_S > I_{NS}$) indicating anisotropy or an orientational order of the clay platelets.

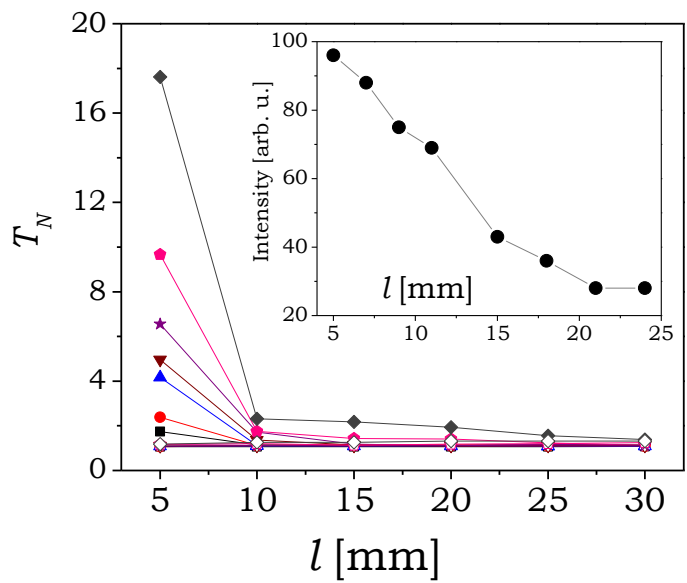

**Figure 2:** Normalized transmittance plotted as a function of distance from the interface along the length of the cuvette for a 2.8 weight % Laponite suspension without externally added salt at different times since preparation (Day 2: squares, Day 4: circles, Day 6: up triangles, Day 8: down triangles, Day 10: stars, Day 14: pentagons and Day 18: diamonds). The filled symbols are for sample with air interface while open symbols

denote the samples with paraffin oil interface. Inset shows the intensity measured through cross polarizer for a sample having air interface using a He- Ne laser for 5mM NaCl suspension 9 days after preparation

We measured the transmittance of the sample through crossed polarizers with air and oil interfaces as a function of distance from the free interface. The measurements could be taken only at a distance of every 5 mm as the optical beam width was large in the microscope even with the optical aperture kept to a minimum. We have plotted normalized transmittance measured at various locations away from the interface for suspension samples having air-Laponite and oil-Laponite interface at various ages in figure 2. It can be seen that the normalized transmittance is enormously increased if the light beam's path is closer to the interface for samples having air interface. The normalized transmittance rapidly decreases for paths more distant from the interface on length-scales of a few millimetres and becomes almost constant in the bulk. Interestingly, the normalized transmittance increases with the age of the sample, much more strongly near the interface than in the bulk. On the other hand, unlike samples with an air interface, transmittance of beams with paths closer to the oil - Laponite suspension interface is observed to be same as the bulk value, which is also equal to transmittance in the bulk of sample with an air interface. We also carried out similar experiments using a He-Ne laser. In these experiments, the laser beam width was about 0.6 mm. Inset in figure 2 shows intensity of transmitted light through a 2.8 wt. % 5 mM NaCl sample having an air interface 9 days after preparation. It can be seen that the transmittance decays rapidly away from the interface and plateaus out in the bulk (about 20 mm from the interface).

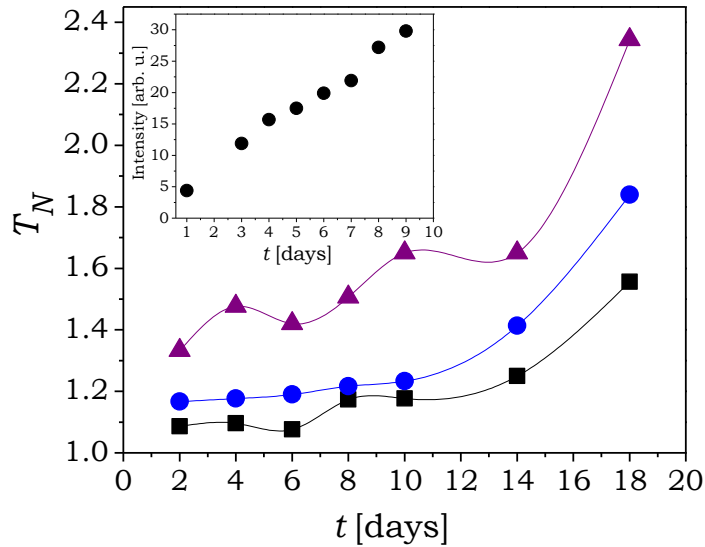

**Figure 3**: Normalized transmittance 25 mm away from the interface plotted as a function of time elapsed since preparation of 2.8 wt. % Laponite suspension for various salt concentrations (squares: 0.1 mM, circles: 3 mM and triangles: 5 mM NaCl). The lines are only guides to the eye (spline fits). Inset shows He-Ne laser intensity measured through cross polarizers for a suspension containing 5 mM salt at the same location.

Further in Figure 3, we have plotted the normalized transmittance through the bulk of the samples (25 mm away from the interface) as a function of time elapsed since sample preparation. It can be seen that the transmittance under the crossed polarizers increases with the age of the samples. In addition, for the samples containing higher concentration of salt, the normalized transmittance is always observed to be higher. The inset in figure 3 shows the intensity of transmitted light from a He-Ne laser at 632 nm wavelength through a sample with 2.8 weight % 5mM NaCl (and measured 25mm away from the air interface). It is clear that the transmittance increases with time elapsed after the sample preparation. We did not observe any noticeable differences in the transmittance for samples having an air interface and an oil interface when the optical path was about 25 mm away from the interfaces.

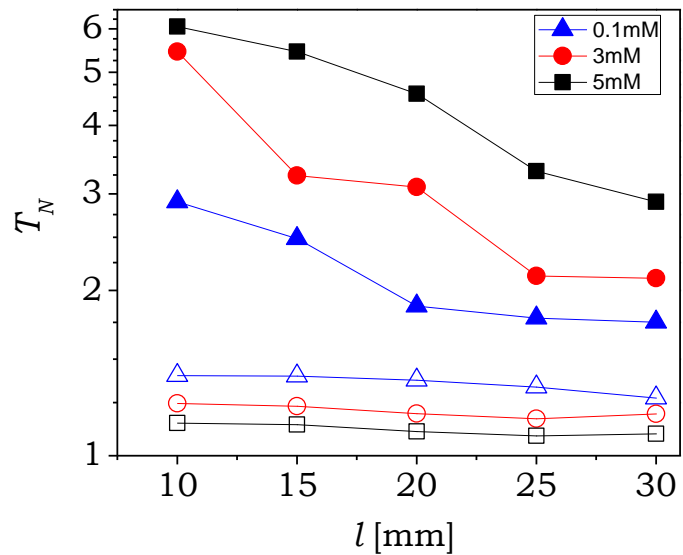

**Figure 4:** Normalized transmittance along the length of a cuvette away from the interface for 2.8 wt. % Laponite suspensions having different concentrations of salt (squares: 0.1 mM, circles: 3 mM and triangles: 5 mM NaCl, line is guide to an eye) 2 days (open symbols) and 21 days (filled symbols) after preparation of the sample.

To understand the effect of salt concentration and age of the sample on the enhancement of optical transmittance, we have plotted the integrated optical transmittance on day 2 and day 21 along the length of the cuvette for 0.1, 3 and 5mM salt concentration Laponite suspensions having an air interface in figure 4. On day 2, although birefringence is observed in the sample, which increases with concentration of salt, no noticeable increase is observed near the interface. On day 21, however, the interface can be seen to be playing a dominating role even beyond 10 mm towards the bulk but keeping within the trend of greater birefringence for higher concentration of salt.

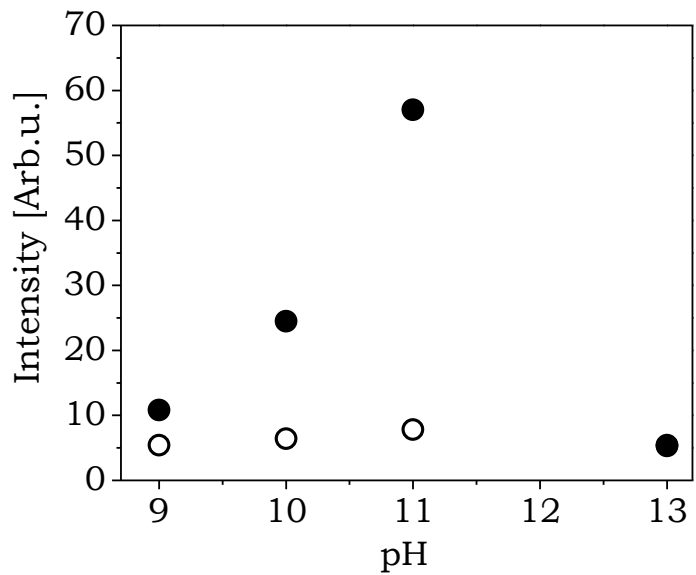

**Figure 5.** Effect of pH on birefringence behavior of 6 days old 2.8 weight % Laponite suspension having 5 mM NaCl. Open symbols represent behavior in bulk (25 mm away from the interface), while filled symbols represent behavior near the interface (5 mm away from the interface).

In addition to salt concentration, we also studied effect of pH of the suspension, concentration of Laponite and temperature at which sample undergoes aging. For 6 days old 2.8 weight % Laponite suspension with 5 mM NaCl having different pH, the intensity of light through the crossed polarizers 5 mm and 25 mm away from the interface is described in figure 5. We observe huge increase in intensity near the interface with increase in pH of suspension from 9 to 11. The enhancement near the interface is observed to be significantly stronger than that of in the bulk. Interestingly at pH 13, no significant enhancement in birefringence either in the bulk or near the interface is observed.

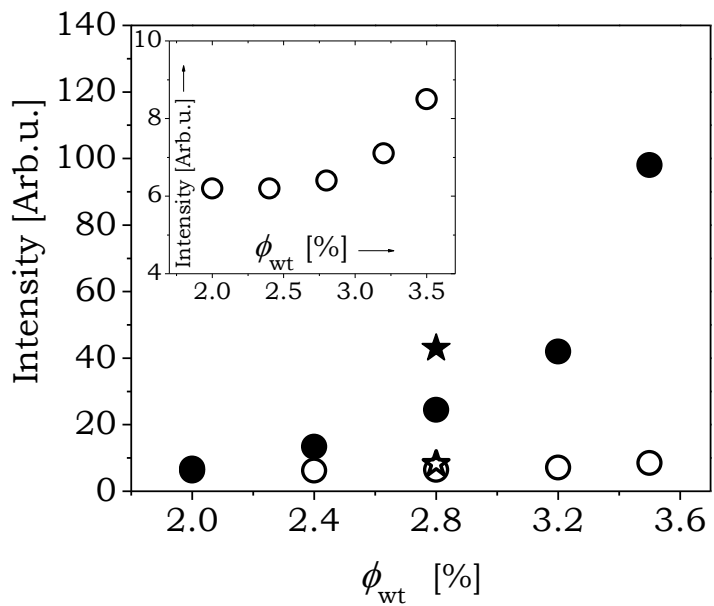

**Figure 6.** Effect of concentration of Laponite on birefringence behavior of 6 days old suspension with pH 10 and 5 mM NaCl. Open symbols represent behavior in bulk (25 mm away from the interface), while filled symbols represent behavior near the interface (5 mm away from the interface). Circles represent experiments carried out on sample aged at 25°C while stars represent that of at 60°C. Inset shows bulk behavior as a function of concentration at 25°C.

Birefringence behavior for 6 days old Laponite suspension having various concentrations (2, 2.4, 2.8 and 3.5 weight %) and 5 mM NaCl near and away from the interface is described in figure 6. It can be seen that for 2 weight % samples no evident difference in birefringence in the bulk as well as near the interface. On the other hand, for higher concentration samples, enhancement in birefringence near the interface is far more pronounced compared to that of in the bulk. We also studied birefringence behavior of a 2.8 weight % 5 mM NaCl sample with pH 10 aged at 60°C. Figure 6 shows that increase in temperature enhances intensity of transmitted light more near the interface than in the bulk.

In order to investigate the orientation of the anisotropy axes for the different interfaces, we measured the transmitted intensity as a function of the angle between the cuvette long axis and the input polarizer. This was accomplished by rotating both the polariser and analyser by the same angle simultaneously while keeping the cuvette fixed. This ensured that the nonhomogeneity of the sample did not affect the measurements as the beam passed through exactly the same region of the sample. Figure 7 shows that the transmitted intensity is almost sinusoidal. This observation confirms that we have a uniaxial system where the transmitted intensity is given by $I_T = I_0 \sin^2(2\alpha)\sin^2(\Delta n k_0 \delta/2)$, where $\alpha$ is angle between the ordering axis in the sample and polarisation axis. The lines in the plot show fits of an expression: $I_T = I_{noise} + \tilde{I}_0 \sin^2(2\theta + \delta)$ to the experimental data. For both the fits $\delta$ is observed to be smaller than 1°. The maximum intensity is observed when the electric field is at an inclination of almost 45° of long side of cuvette indicating that within the accuracy of the measurements the anisotropy axes match with the cuvette long axis and the interface. The modulation of the transmittance with the angle was also measured at various distances from the air interface. The figure shows that at about 2 mm from the interface magnitude of transmittance that varies sinusoidally is significant. The intensity decreases with increase in distance from interface. Beyond a distance of 30 mm from air interface the modulation becomes minimal so that the sinusoidal variation cannot be clearly observed because of noise and the inhomogeneous anisotropy of the sample. The transmittance in the oil interface samples is not significant as compared to the bulk and very closely matches the transmittance measured at a distance of 30 mm from the air interface. The transmittance variation indicates a global anisotropy that stems from a long range directional ordering within the Laponite suspension. This presumably arises from the orientational ordering of the meso-structures in the jammed Laponite suspension. Interestingly the angle does not depend on the salt concentration indicating that only the interface may have a role. The angular dependence sets in the sample only after a period of 8 days.

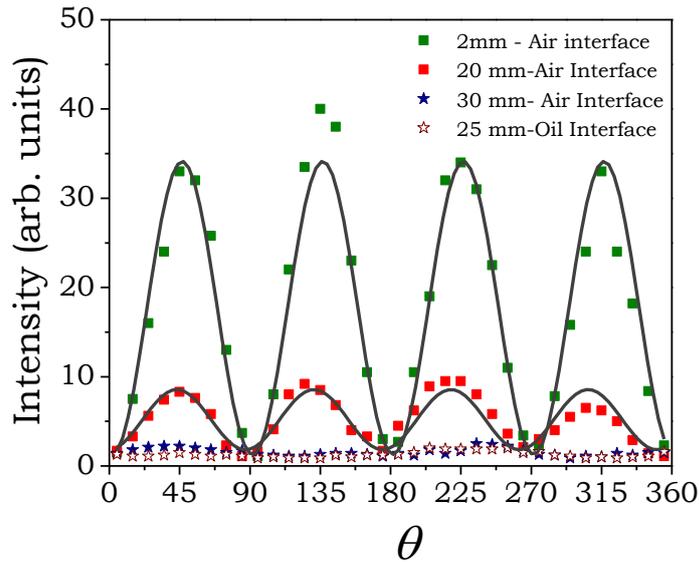

**Figure 7**: Intensity variation for a 2.8 weight % 5mM salt system 12 days after preparation plotted as a function of *θ*, the angle between the cuvette long axis and the input polarizer, obtained by rotating simultaneously the polariser and analyser with respect to the sample. The variations obtained along the length of the cuvette for samples with air and oil interface are shown.

## IV. Discussion

We have investigated the optical birefringence in jammed aqueous Laponite suspensions with air and oil interfaces. The various observations reported in this study can be broadly divided into four issues. 1) The observation of anisotropic orientation and how it is influenced by aging time, concentration of salt, concentration of Laponite, pH of suspension and temperature. 2) An increase in birefringence while system is in a physically arrested state. 3) Effect of an interface, air and/or paraffin oil on the microstructure. 4) The observation of interface induced anisotropy over length-scales of several orders of magnitude greater than the particle length-scale.

In aqueous suspensions of Laponite, the disk like shape and the interplay between attraction and repulsion among the particles influence the morphology and phase behaviour. Many groups have proposed that ergodicity breaking in aqueous suspension of Laponite beyond 1 volume % is due to repulsive interactions among Laponite particles that leads to the formation of a Wigner glass.[8, 12, 14, 42] Orientational order in such a situation, where Laponite particles are surrounded by neighboring particles due to repulsive interactions (without physically touching each other), can be analyzed by Onsager's theory for suspension of hard disks. The reduction in excluded volume by aligning the particles drives the isotropic - nematic transition in anisotropic systems.[43] Suspension of hard discs, for an aspect ratio $a$ (= $L/D$) is expected to demonstrate an isotropic to biphasic and biphasic to nematic transition above the volume fractions $\phi_{i-b}$=0.33$a$ and $\phi_{b-n}$=0.45$a$ respectively. In biphasic state, both nematic and isotropic phases co-exist. For Laponite RD, whose aspect ratio is in a range: 25 - 30, a concentration at which isotropic to biphasic transition should be observed is 2.75 weight % while a nematic state is possible only above 3.8 weight %.[10] However, in order to have a self-suspended state of Laponite particles originating from repulsive interactions, for the concentration of 2.8 weight %, Laponite particles need to maintain an average distance of 40 nm from each other.[10, 13] A recent small angle X-Ray scattering study by Ruzicka and coworkers[12] carried out for similar concentrations of Laponite without any externally added salt aged for around 50 hours reports a peak in structure factor at $Q$ =0.17nm$^{-1}$ corresponding to an interparticle distance of 37 nm. Since Laponite has a diameter of 25-30 nm, the jammed state is proposed to be a repulsive glass. Ruzicka and coworkers[12] also carried out theoretical and simulation studies with Debye screening length in the range 5 to 10 nm that showed agreement with their experimental study by considering Wigner (repulsive) glass picture.

**Table 1.** Conductivity and Debye screening length of 18 days old 2.8 weight % Laponite suspension

| Concentration of salt (mM) | Conductivity (μS/cm) | Debye screening length (nm) |
|---|---|---|
| 0.1 | 956 | 3.1 |
| 3 | 1250 | 2.8 |
| 5 | 1380 | 2.71 |

Interestingly we observe that birefringence associated with samples having higher concentrations of salt was always greater at any aging time. In order to judge this behavior from a point of view of repulsive glass scenario, information on Debye screening length associated with the same is necessary, which requires concentration of cations (Na$^+$ ions) and anions (Cl$^-$ ions) in the suspension. Concentration of ions can be obtained from conductivity of Laponite suspension having varying concentration of salt and is given by: $\sigma = e(\mu_+ n_+ + \mu_- n_-)$,[44] where $\sigma$ is conductivity, $\mu_+$ and $\mu_-$ are mobilities of cations and anions respectively, $n_+$ and $n_-$ are concentrations of

cations – and anions respectively and *e* is the electron charge. Presence of Cl⁻ ions in the suspension is due to only externally added salt while Na⁺ ions are present because of externally added salt as well as counterions. Ignoring the concentration of OH⁻ ions,[13] $n_-$ can be directly obtained from concentration of externally added salt. Therefore, with knowledge of mobilities of Na⁺ and Cl⁻ ions ($\mu_{Na}$ =5.19×10⁻⁸ m²/sV and $\mu_{Cl}$ =7.91×10⁻⁸ m²/sV)[45] and conductivity of Laponite suspension, $n_+$ can be easily estimated, which in turn can be used to estimate Debye screening length. In table 1, we report the ionic conductivity ($\sigma$) of 2.8 weight % aqueous Laponite suspension, 18 days after preparation of suspension along with Debye screening length for 0.1, 3 and 5 mM salt concentration. It can be seen that increase in concentration of salt causes decrease in the Debye screening length. This observation suggests decrease in repulsive interactions with increase in the concentration of salt. Figures 3 and 4 show that the enhancement in birefringence is faster with increase in salt concentration. As discussed in the introduction, enhancement of viscosity is also faster with increase in salt concentration.[10, 25] If we believe that enhancement in viscosity and birefringence in 2.8 weight % Laponite suspensions with varying salt concentration is due to same microstructure, repulsive interaction scenario cannot explain faster build-up of Wigner glass at higher salt concentration. Interestingly enhancement of viscosity and birefringence seem to have significant correlation. As reported in the literature, increase in concentration of Laponite as well as temperature causes faster increase in viscosity of its aqueous suspension.[25, 46] Similarly we also observe greater enhancement in birefringence with increase in concentration and temperature as shown in figure 6.

The other popular proposal for a microstructure of aqueous Laponite suspension is an edge – to – surface attractive interaction originating from positively charged edge and negatively charged face leading to a gel. The most popular attractive configuration is a house of cards structure where there is an edge-to-surface attractive interaction among the Laponite particles as shown in figure 8(a) which produces a gel like structure.[15, 47-50] Recently Jonsson *et al.*[19] proposed another possibility wherein edge-to-surface association occurs in an overlapped coin configuration (figure 8(b)). Their Monte Carlo calculations predict that for ionic concentrations in the range of 10 to 60 mM, overlapped coin configuration has a deeper minimum compared to perpendicular edge-to-face configuration. In order to have two particles approach each other to have an edge-to-surface interaction, they need to overcome a repulsive barrier between them. With increase in concentration of salt, repulsion among particles reduces, decreasing an energy barrier for particles to approach each other. Interestingly increase in viscosity is known to be get expedited in presence of salt.[10, 25] Since phenomenon of enhancement in viscosity can arise from formation edge-to-surface bonds, this observation suggests possibility of co-relation between orientational anisotropy and attractive associations in Laponite suspension. As shown in figure 5, Laponite suspension shows greater birefringence with increase in pH up to 11, however does not show any noticeable birefringence at pH 13. As suggested in the introduction section, at pH 13, edge of Laponite particle possesses a negative charge, therefore edge – to – face interactions are certainly not possible at that pH. On the other hand with increase in pH from 9 to 11, greater ionic strength of the suspension (due to greater concentration of NaOH in high pH system) may lead to greater birefringence.

Overall, the repulsive glass scenario does not explain the salt concentration dependence of order formation, while the attractive gel scenario does not explain the interparticle distance observed experimentally. We therefore feel that the issue of microstructure of aqueous Laponite suspension is still open and further spectroscopic studies such as SAXS are needed to be conducted at different concentrations of salt and pH over a greater span of aging times.

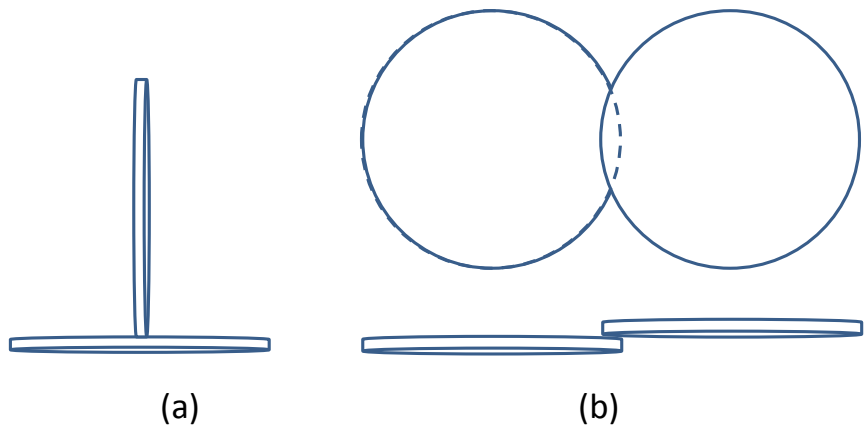

(a)   (b)

**Figure 8.** Schematic of possible edge – to - surface attractive interactions. (a) Edge – to - surface perpendicular interaction (T shaped) leading to a house of cards structure. (b) Edge - to - surface overlapping coin configuration proposed by Jonsson *et al.*[19]

It is usually observed that viscosity of 2.8 weight % Laponite suspension increases very rapidly after mixing Laponite with water such that it supports its own weight within few hours of sample preparation. However, as shown in figures 2 and 3, it takes few days for the suspension to show significant birefringence. This behaviour suggests that the motion of the particles in a jammed state is not completely frozen and undergoes local orientational adjustments which keep evolving as a function of time. Eventually the clay platelets form micro-domains with the director of each domain oriented in a particular direction. It is difficult to comment on how this process occurs unless the precise cause of such massive increase in viscosity is known. If we assume that the repulsive interactions cause the increase in viscosity; then in a jammed state, order formation can take place by movement of individual Laponite particles. If edges – to – surface interactions are favoured, two possibilities exist as described in figure 8(a) and (b). According to Jonsson and coworkers,[19] an overlapped coin configuration has a deeper but much narrower free energy minima while T configuration has shallower but broad minima. Therefore, there is a possibility that in a jammed state, edge – to – surface interaction may undergo a transition from T configuration to overlapped coin configuration during an aging process that not only increases its viscosity further, but also induces an orientational order. Owing to restricted mobility of Laponite particles in a jammed state such restructuring is expected to be very slow in the bulk of the material as observed in figures 2 and 3.

Now we turn to the effect of the interface on the morphology of Laponite suspensions. We observe an alignment of Laponite particles near an air interface, which otherwise is absent in a sample having an interface with paraffin oil. Figure 7 suggests that Laponite particles are oriented almost perpendicular to the interface (surface normal of a disc is parallel to the interface). It is apparent from figure 2 that birefringence increases with a sample age much more rapidly at the interface than in the bulk. Thermodynamically, migration of Laponite particles to the air interface will be favourable, if it causes decrease in the surface energy of air-water interface. The surface energy (interfacial tension) of water - air interface is 73 mJ/m$^2$,[51] while that of paraffin oil interface is lower at 48 mJ/m$^2$,[52] Although there is no report in the literature about the effect of Laponite on the surface energy of air - water or air - oil interface; Na-Montmorillonite, which has similar crystal structure to that of Laponite, is reported to increase the surface energy of water - air interface.[53] Owing to the similarity of the structure,[2] if we assume that Laponite also increases surface energy of water-air interface, migration of Laponite particles towards the interface will not be favoured. Furthermore, such scenario will repel Laponite from water-air interface creating a thin layer near the air interface which is depleted of Laponite particles.[54] Surface energy drive, therefore, should be ruled out as a reason behind such order formation.

The other possibility discussed in the literature which may lead to an understanding of the observed phenomenon is the greater mobility of the particles near the interface. It has been proposed that in glassy materials, particles near a free surface experience a structural cage only from the inner side, which renders them with better mobility than the bulk and may expedite aging.[55] It is important to note that in the bulk of the Laponite suspension, Laponite particles are surrounded by neighbouring particles that significantly hinder their mobility taking the system into an arrested state.[56] However, since the particles near the air interface are depleted from a very thin layer near the interface, their mobility is expected to be enhanced from one side. Figure 3 suggests that aging in 2.8 weight % Laponite suspension involves setting in an order as a function of time. Therefore it is possible that the inherent tendency of Laponite particles to form an ordered state (specific microstructure that leads to ordering is still an open question) gets amplified due to greater mobility of the particles near the free surface. This phenomenon may drive nematic ordering of the particles at the air interface, which otherwise is difficult in the bulk due to hindrance from all the sides. Recently Mamane and coworkers [57] studied thermal fluctuations of the free surface of aging Laponite suspension and observed that older Laponite suspension samples exhibit bursts of fluctuations that demonstrate non-Gaussian dynamics compared to younger Laponite suspensions. Concentration of Laponite used by Mamane and coworkers (2.5 wt. %) is not very different from that used in the present study (2.8 wt. %). They also suggested that the observed behaviour could result from enhanced mobility of the Laponite particles near the free surface. The perplexing behaviour, however, is that the greater ordering near the interface than in the bulk is absent in case of paraffin oil interface, as the physical hindrance towards the interface will not be present in this situation as well. There may be a possibility that oil specific interactions (or absence of interactions) of Laponite due to its organophobic nature may be responsible for prevention of the order formation. Nonetheless it appears from the observed behavior that the faster structure (gel or glass) formation near the air interface that progressively percolates towards the bulk is responsible for the observed behavior. Changes in the interaction potential around the particles near air interface may also drive such phenomena. Absence of structure formation in oil interface system, reason of which is not entirely clear to us at this moment, does not lead to birefringence near the interface. In order to have better understanding of this behaviour, SAXS studies are needed to be carried out in the vicinity of air or oil interface.

An astonishing result of the present work is that the influence of order in samples having air interface is observed beyond five orders of magnitude greater length-scales than the particle diameter (particle diameter is 30 nm while order is observed beyond 5 mm) as well as the average inter-particle distance. Furthermore, the depth over which interface seems to be influencing is observed to increase with presence of salt. Stevenson and Wolynes[55] suggested that enhanced mobility at the surface percolates into the bulk through collaborative dynamics. The depth up to which the effect is observed would then be determined by length scale of the cooperativity. Considering a recent report that disorder on the surface destroys long-range order in the bulk (percolation of the surface phenomenon to the bulk),[58] the present observation, which demonstrates percolation of order to the bulk, may not be surprising. Again, if edge – to - face interactions are existent, formation of energetically preferred overlapped coin configuration near the interface may trigger formation of same microstructure into the bulk of the material like a zipping phenomenon.[59]

Finally, we concur that our results raise more questions about the phase behaviour of aqueous Laponite suspension than the answers. We believe that spectroscopic investigations near the interface and realistic simulations on charged disks may lead to further insight into the observed phenomena.

**V. Conclusion:**

We study birefringence behaviour of aqueous suspension of Laponite for samples having air and paraffin oil interface. We observe that for both the types of interfaces, the suspension is weakly birefringent in the bulk. In case of air interface, however, the suspension shows much larger birefringence near the interface suggesting a nematic order whose effect percolates beyond five orders of magnitude greater length-scale than the particle length-scale below the surface. No such behaviour is observed for Laponite suspension with an oil interface. In addition, the order increases with time even though the suspension is in a physically jammed state. Interestingly at any aging time as well as location from the interface, greater order is observed for greater concentration of salt and greater concentration of Laponite in the suspension.

We believe that, owing to its anisotropic shape, Laponite particles in aqueous suspension with 2.8 weight % concentration have an inherent tendency to set in order. On the other hand since the structural arrest precedes order formation, system demonstrates weak order in the bulk. It appears that anisotropic structure formation, which leads to ergodicity breaking in the suspension, originates near an air interface. We believe that the cooperative dynamics of particles near the interface are responsible for a prorogation of the surface phenomenon into bulk over the length-scales that are several orders of magnitude greater than the particle dimensions. Further spectroscopic investigations to establish the structure and the ordering mechanism are required.

**Acknowledgement**: This work was supported by the Department of Science and Technology, Government of India under IRHPA scheme.